# The effect of grain boundaries on magnetic exchange interactions in iron

Martin Zelený[1*], Martin Heczko[2], Petr Šesták[3,4], Denis Ledue[5], Renaud Patte[5] and Miroslav Černý[3,4]

[1]Institute of Materials Science and Engineering, Faculty of Mechanical Engineering, Brno University of Technology, Technická 2, CZ-61669 Brno, Czech Republic

[2]Faculty of Metals Engineering and Industrial Computer Science, AGH University of Krakow, Czarnowiejska 66, PL-30054 Kraków, Poland

[3]Institute of Physical Engineering, Faculty of Mechanical Engineering, Brno University of Technology, Technická 2, CZ-61669 Brno, Czech Republic

[4]Central European Institute of Technology (CEITEC), Brno University of Technology, Purkyňova 123, CZ-61200 Brno, Czech Republic

[5]Normandie Université, UNIROUEN, INSA Rouen, CNRS, GPM, F-76800 Saint Étienne du Rouvray, France

*zeleny@fme.vutbr.cz

**Abstract**

This work investigates how grain boundaries (GBs) modify magnetic exchange interactions in bcc iron, with particular focus on the effect of phosphorus segregation. Using density-functional theory combined with the Liechtenstein–Katsnelson–Antropov–Gubanov Green's-function approach, we calculate Heisenberg exchange parameters for three symmetric tilt GBs, Σ5(310), Σ13(510), and Σ13(320), and use these parameters in Monte Carlo simulations to evaluate finite-temperature magnetic behavior. All clean GBs exhibit strong local deviations from bulk exchange interactions, including antiferromagnetic coupling across the boundary plane. These negative exchange interactions are not governed by interatomic distance alone, but arise primarily from the altered local coordination and symmetry breaking at the GB. Phosphorus segregation, modeled in both substitutional and interstitial configurations at the Σ5(310) GB, suppresses the antiferromagnetic couplings and significantly redistributes the local exchange landscape through chemical and electronic effects. Monte Carlo results show that, despite pronounced local perturbations, realistic GB densities cause only a small reduction in the Curie temperature because bulk-like regions dominate the global magnetic transition. A substantial decrease in Curie temperature appears only when the GB volume fraction is artificially increased. The results demonstrate that GBs strongly influence local magnetic interactions while having a limited effect on global magnetic ordering, and they establish a general framework for linking atomistic interfacial structure and chemistry to mesoscale magnetic behavior in Fe-based materials.

**Introduction**

Lattice defects play an essential role in determining many properties of crystalline materials. Point defects, such as vacancies and interstitials, are primarily responsible for influencing thermal and electrical properties, whereas the mechanical behavior of materials is controlled by extended defects, including dislocations and planar defects like grain boundaries (GBs) and phase boundaries. Magnetic properties, in particular, are highly sensitive to the presence of lattice defects [1]. Impurity atoms in the crystal lattice can locally alter the electronic structure, leading to the suppression or enhancement of magnetic moments and spin exchange interactions in their vicinity [2,3]. Additionally, GBs and phase boundaries serve as obstacles to the movement of magnetic domain walls [4,5], often pinning them [6] and thus impacting macroscopic magnetic properties such as coercivity, remanence, and magnetic saturation. The density and distribution of planar defects within a material significantly affect its magnetic performance. Grain boundaries can also modify magnetic anisotropy due to changes in the preferred magnetization axis at the boundary [7,8], potentially introducing additional magnetic domains. Furthermore, impurity atoms in the host lattice may segregate at GBs, creating localized regions with dramatically different chemical environments [9,10]. This segregation can lead to the formation of magnetically "dead layers," where the magnetic properties are suppressed or entirely diminished [11-13]. Therefore, in polycrystalline materials, controlled heat treatment can adjust grain size and GB density to tune magnetic response. In advanced soft magnets, nanoscale GB engineering—promoting stable ultrafine grains, a favorable boundary character distribution, and tailored GB chemistry—delivers high permeability and low coercivity while maintaining thermal stability [14,15].

Despite the critical role of GBs in influencing magnetic properties, a fundamental gap remains in how atomic-scale GB structure controls exchange interactions. The exchange interactions between spins at sites *i* and *j*, quantified by $J_{ij}$ in the Heisenberg Hamiltonian

$$H = -\sum_{\langle i,j\rangle} J_{ij} \vec{S}_i \cdot \vec{S}_j, \tag{1}$$

(with unit-length spins $|\vec{S}_i| = 1$; under this sign convention $J_{ij} > 0$ favors ferromagnetic alignment), provides the bridge from electronic structure to magnetic phase transitions [16]. Once the set of $J_{ij}$ is determined, finite-temperature equilibrium behavior can be modeled using Monte Carlo (MC) or atomistic spin-dynamics simulations. A widely used way to obtain $J_{ij}$ is to map first-principles energies of (constrained) magnetic configurations in a supercell onto a Heisenberg model; at GBs this mapping can be delicate because structural disorder, long-range relaxations, and supercell geometry affect the fitted parameters [17]. Alternatively, the Green's-function linear-response approach of Liechtenstein–Katsnelson–Antropov–Gubanov (LKAG) evaluates $J_{ij}$ directly from the electronic structure via the magnetic-force theorem by treating infinitesimal local spin rotations as a perturbation [16]. The $J_{ij}$ parameters are well documented for simple metals and for their interfaces [18-20], yet much less is known about their behavior near GBs. At GBs, atomic-level irregularity alters interatomic distances and angles, which can renormalize—or even invert—the sign of exchange, consistent with the classic Goodenough–Kanamori–Anderson framework for angle-dependent exchange pathways [21-23]. First-principles studies of representative GBs already show strong local-moment variations at the boundary, underscoring the expectation of modified $J_{ij}$ and possible spin frustration in GB regions [24].

Various approximations have been proposed to account for the contribution of GBs in modeling magnetic systems. These methods often simplify the complex atomic structure at GBs by assuming average local environments or by modifying exchange-interaction or anisotropy parameters to emulate local disorder. In some cases, mean-field-type treatments or random-anisotropy–style averaging are used to estimate the impact of GBs, and fitting simulation parameters (e.g., anisotropy constants) to experimental data is common practice [8, 25]. However, such approaches can overlook atomic-scale variations and miss some important phenomena—such as spin frustration and local magnetic inhomogeneities—that arise from the irregular atomic arrangement at GBs.

To address this gap, this study examines how the atomic structure in the vicinity of symmetric tilt GBs in iron influences the exchange-interaction parameters $J_{ij}$, with particular emphasis on the effect of phosphorus (P) segregation. Using *ab initio* density-functional theory (DFT), we determine $J_{ij}$ for a range of GB configurations and then employ the importance sampling MC method to model equilibrium magnetic behavior at finite temperatures, enabling assessment of temperature dependence of the magnetization, susceptibility and heat capacity and consequently Curie temperature. The primary aim is to develop and demonstrate a reliable method for estimating $J_{ij}$ at GBs using iron as a model system, with the intent to extend the framework to more complex, application-oriented materials (e.g., soft magnets and permanent magnets) where microstructural control is central to performance optimization. We additionally investigate P segregation at iron GBs: phosphorus is a common impurity in Fe-based alloys, known to segregate strongly to ferrite GBs and to influence cohesion/embrittlement, making it a useful probe of how a nonmagnetic p-element impurity modifies exchange interactions via electronic and chemical perturbations [26-29]. These insights clarify how local atomic structure and chemical variation at GBs impact $J_{ij}$ and magnetic properties at finite temperatures, and they establish a general methodological framework applicable across materials classes.

**Computational details**

The exchange interaction parameters $J_{ij}$ were calculated based on collinear ferromagnetic configurations using the Green's function method according to the LKAG formula [16] implemented in the TB2J package [30]. *Ab initio* calculations required for estimation of $J_{ij}$ parameters have been performed within the framework of non-local density functional theory as implemented in the Siesta (Spanish Initiative for Electronic Simulations with Thousands of Atoms) package [31] Max-1.2.0 version. The core electrons were described by norm-conserving scalar relativistic pseudopotentials with non-linear core corrections [32,33]. The valence electron wave functions were expanded over a double-ζ polarized basis set of finite-range numerical pseudoatomic orbitals [34,35]. The localization radii of the basis functions were determined from an energy shift of 0.01 Ry and the energy cutoff for real-space mesh size was set to 300 Ry. The exchange-correlation functional was described within the Perdew–Burke–Ernzerhof (PBE) parametrization of generalized gradient approximation [36]. The combination of numerical pseudoatomic orbitals together with norm-conserving pseudopotentials has already been successfully applied to the description of the magnetic properties and GBs in Fe [37,38]. Using the localized basis set allows direct construction of the tight-binding model for Green's function method necessary for estimation of $J_{ij}$ parameters. It makes this approach advantageous compared to employment of a plane-wave basis set which requires an additional step comprising projection of delocalized wave-function onto orbitals described by Wannier functions. With these calculation settings and 21 × 21 × 21 k-point sampling grid of the Brillouin zone, we obtained equilibrium lattice parameter of bcc Fe $a_{\mathrm{Fe}}$ = 5.39 a.u. and magnetic exchange-interactions parameters for the nearest neighbors and the next-nearest neighbors $J_{NN}$ and $J_{NNN}$ equal to 1.14 mRy and 0.78 mRy, respectively. These results correspond very well to previously published data [16,39-41].

The GBs were modeled by supercells containing two equivalent, reversely oriented GB planes constructed according to the coincidence site lattice (CSL) theory by rotating two bcc grains around the [001] axis [42]. In particular we created supercells for three different GBs in Fe, namely Σ5(310) with 120 atoms, Σ13(320) with 94 atoms and Σ13(510) with 152 atoms (see Figs. 1c and 4c-f). The corresponding k-point sampling grids were 6 × 1 × 13, 6 × 1 × 13, and 5 × 2 × 11, respectively for *ab initio* calculations and structural optimization with Siesta. The denser grids 9 × 1 × 15, 9 × 3 × 15, and 5 × 2 × 13 were used for $J_{ij}$ calculations with TB2J. To study the effect of P impurities at Σ5(310) GB we included two P atoms at the interstitial positions exhibiting the largest space at GB plane. Since we are particularly interested in exchange interactions across the GB, we also considered substitutional P atoms replacing all Fe atoms in the GB plane in order to preserve boundary symmetry, although a previous study [43] found more energetically favorable segregation sites in the neighboring planes. All atomic positions and supercells size in the direction perpendicular to the GB plane were optimized using the combined ionic-unit cell dynamics [44] with Broyden–Fletcher–Goldfarb–Shanno quasi-Newton algorithm implemented in Atomic Simulation Environment (ASE) [45] until the force on each atom was less than 0.4 mRy/a.u.

MC simulations were based on the classical Heisenberg Hamiltonian (eq. (1)) and carried out only for Σ5(310) GB. The original supercell was expanded to a 4 × 2 × 10 supercell containing 9600 atoms. Only $J_{ij}$ interaction parameters for nearest and next-nearest neighbors up to a distance 7.2 a.u. were considered. To investigate the magnetic phase transition at finite temperature, we used the importance sampling MC method [46,47] based on the standard Metropolis algorithm [48]. At each MC step (MCS), a new possible orientation of each spin is randomly chosen in all directions. At each temperature, $5\times10^4$ MCS were performed, the first $10^4$ MCS were discarded before averaging over the last $4\times10^4$ MCS to calculate the thermodynamic quantities (internal energy, heat capacity, spontaneous magnetization and susceptibility). In order to reduce statistical errors, the final estimates of the thermodynamic quantities were obtained by averaging over 20 independent simulations performed in parallel, giving 800,000 production MCS at each temperature. Each simulation started from a different initial magnetic configuration and used a different random number sequence. The (reduced) spontaneous magnetization is defined by:

$$M(T) = \frac{1}{N} < \left\| \sum_{i=1}^{N} \vec{S}_i \right\| >_T, \tag{2}$$

where $< >_T$ means statistical (Gibbs) average. In the MC simulations, as mentioned above, this average is calculated by averaging over the MCS thanks to the ergodicity principle. To locate the transition temperature, we calculated the susceptibility $\chi(T)$:

$$\chi(T) = N \frac{<m^2>_T - <m>_T^2}{k_B T} \tag{3}$$

with $m = \frac{1}{N} \left\| \sum_i \vec{S}_i \right\|$ ($\chi$ is proportional to the true linear susceptibility) and the heat capacity per spin *C(T)*:

$$C(T) = \frac{<E^2>_T - <E>_T^2}{N k_B T^2}, \tag{4}$$

where *E* is the magnetic energy of the crystal calculated with eq. (1). In the case of a second order transition, $\chi(T)$ and *C(T)* exhibit a divergence or a discontinuity at the transition temperature $T_c$ in the thermodynamic limit ($N \rightarrow \infty$). In a finite crystal, the curves are rounded because of finite-size effects [47,49,50] and $\chi(T)$ and *C(T)* exhibit a maximum close to $T_c$ if *N* is large enough.

**Results**

To clarify how GB structure affects magnetic exchange in iron, we first analyzed the clean Σ5(310) GB. This system serves as a benchmark for disentangling intrinsic GB effects from impurity-segregation effects. The calculated exchange parameters $J_{ij}$ are shown in Fig. 1a as a function of the *y* coordinate of the pair center (the coordinate normal to the GB plane) measured from the center of the computational cell. Atomic pairs up to an interatomic distance *d* = 7.2 a.u. are included; the corresponding distance distribution is summarized in Fig. 1b. Within ~11 a.u. of the GB plane, $J_{ij}$ deviates markedly from the bulk values for the nearest-neighbor ($J_{NN}$) and next-nearest-neighbor ($J_{NNN}$) shells, which are clearly identifiable in the bulk-like region of the supercell. Exactly at the GB, some $J_{ij}$ become negative, with a minimum of −1.2 mRy, indicating strong antiferromagnetic coupling across the boundary. The specific atomic pairs exhibiting negative $J_{ij}$ are highlighted in Fig. 2a via a color map of exchange strength in the GB vicinity. The strongest negative interaction occurs between atoms in the first layers adjacent to the GB plane, which also have the shortest interatomic distance, *d* = 4.27 a.u. (denoted as (I) in Fig. 2a). A second negative value, $J_{ij}$ = −0.44 mRy, appears for pairs connecting the first and second layers across the GB (II), despite their largest interatomic distance among the evaluated pairs, *d* = 7.21 a.u. By contrast, the interaction between atoms in the second layers across the GB (III) is positive but small ($J_{ij}$ = 0.24 mRy) at *d* = 7.15 a.u. Very weak couplings are also found for the next-nearest interaction between atoms in the first layer and atoms in the GB plane ((IV) $J_{ij}$ = 0.27 mRy), as well as for interactions between atoms lying within the GB plane at a distance equal to the bcc-Fe lattice parameter (not shown in Fig. 2a because this bond is oriented parallel to the viewing direction). Enhanced couplings are also observed near the GB. For example, the nearest-neighbor interaction between atoms in the first layer and atoms in the GB plane (V) reaches $J_{ij}$ = 1.44 mRy, which exceeds the corresponding bulk value of 1.15 mRy. The associated interatomic distance, *d* = 4.82 a.u., is slightly longer than the bulk nearest-neighbor distance $d_{NN}$ = 4.67 a.u.

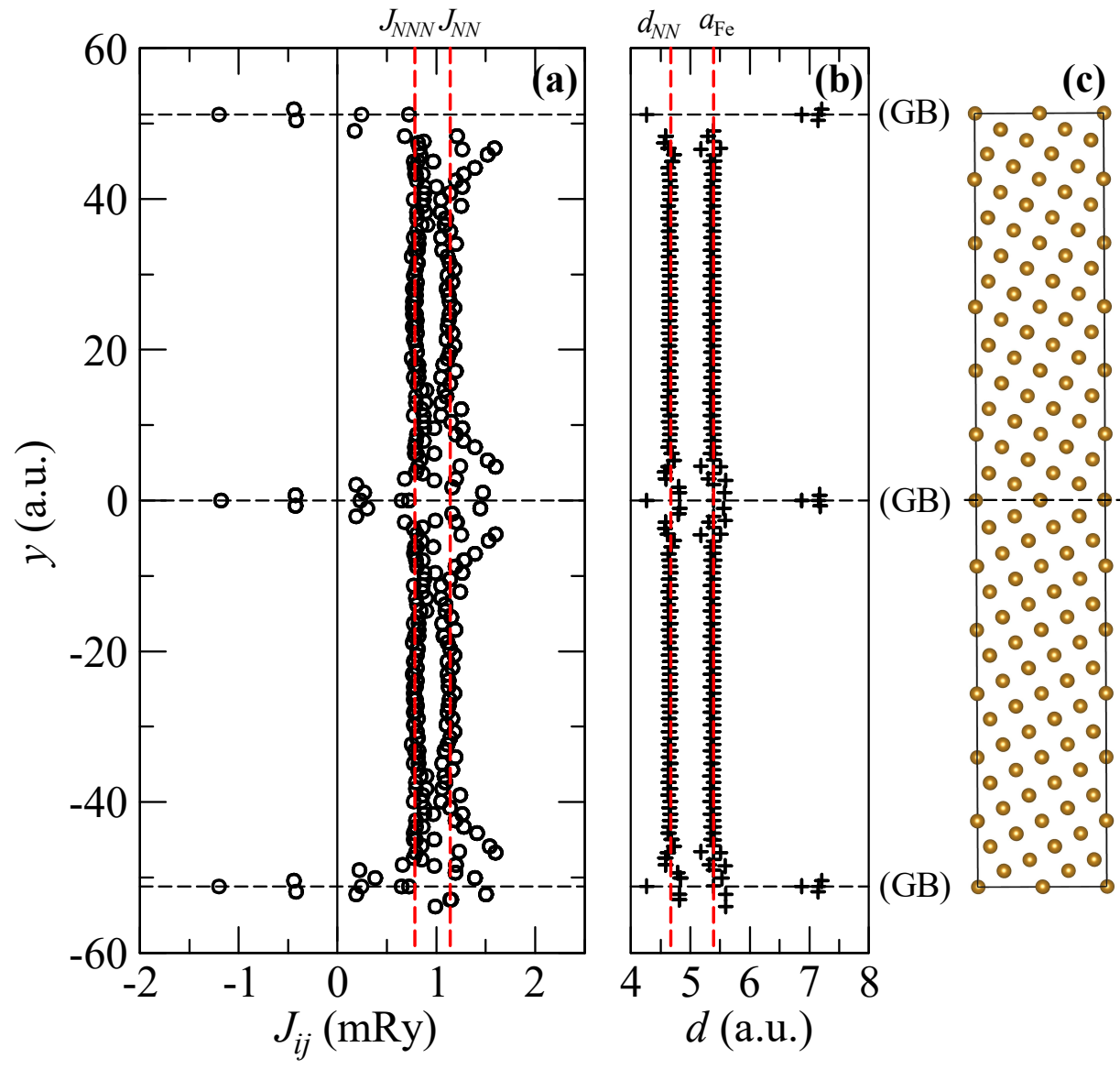

Fig. 1: (a) Exchange interaction parameters $J_{ij}$ and (b) corresponding interatomic distances *d* for the nearest and next-nearest neighbors in clean Σ5(310) GB. The vertical axis on the plots represents the *y*-coordinate of the center of the atomic pair *i-j* with respect to the center of the computational cell. The black dashed horizontal lines mark the GB planes, while the red dashed vertical lines denote the corresponding values of $J_{NN}$, $J_{NNN}$, $d_{NN}$ and $d_{NNN} = a_{\text{Fe}}$ in bulk iron. (c) Computational cell of clean Σ5(310) GB used for ab initio calculations.

After establishing this reference behavior, we investigated the effect of phosphorus impurities at the boundary. Both substitutional and interstitial configurations were considered to evaluate how chemical and structural perturbations modify the local exchange interactions and potentially suppress or enhance magnetic coupling at the interface. Previous theoretical calculations and experimental observations show that P atoms preferentially occupy interstitial sites at the GB [43,51]. We note that the structural models used here contain a relatively high impurity concentration—higher than the experimentally observed maximum solubility of P in Fe (0.033 at.%) [51]—because of the small supercells required under periodic boundary conditions. Nevertheless, even when the bulk solubility is low, impurity atoms can reach significantly elevated local concentrations at grain boundaries, making such configurations physically relevant.

For the Σ5(310) GB with substitutionally segregated P, the exchange-interaction pattern differs markedly from the phosphorus-free boundary (Fig. 2b). The interaction across the GB plane between atoms in the first adjacent layers (I) now exhibits the largest positive value, $J_{ij}$ = 1.93 mRy, with a slightly shorter interatomic distance *d* = 4.19 a.u. compared to the clean GB. Other interactions across the GB plane, (II) and (III), are nearly zero. A negative exchange value, $J_{ij}$ = −0.57 mRy, also appears in the substitutional case, but it occurs within the same grain, between next-nearest neighbors connecting the first and second adjacent layers (VI). Most other interactions near the GB show enhanced nearest-neighbor couplings but suppressed next-nearest-neighbor couplings (Fig. 3a). The region where exchange interactions are affected by P segregation is also broader than in the clean GB, extending up to approximately 16 a.u. from the interface. In contrast, structural distortions—reflected in changes in interatomic distances—remain confined to a much narrower zone around the GB plane (Fig. 3b).

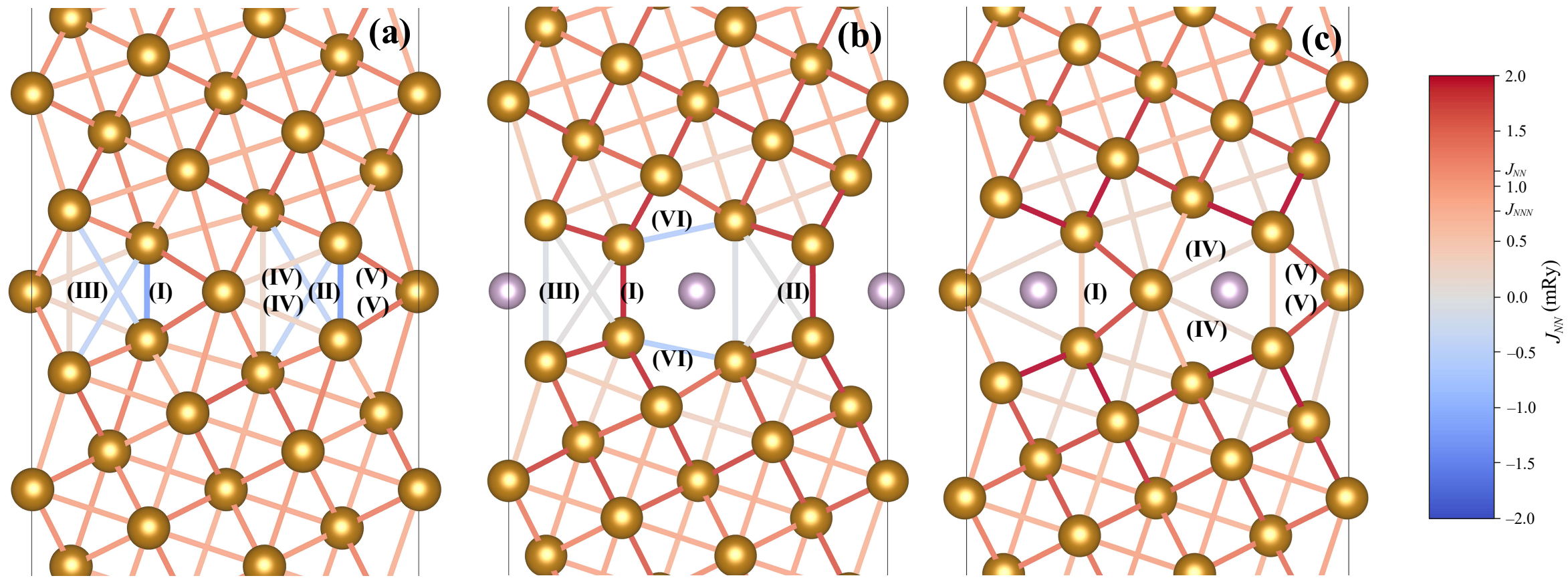

Fig. 2: Color maps of the exchange-interaction strength $J_{ij}$ for (a) the clean Σ5(310) GB, (b) the GB with substitutionally segregated P atoms, and (c) the GB with an interstitially segregated P atom. Red shading indicates ferromagnetic coupling (positive $J_{ij}$), whereas blue shading corresponds to antiferromagnetic coupling (negative $J_{ij}$). Fe atoms are shown in gold and P atoms in purple. Labels (I)–(VI) mark specific atomic pairs whose interactions are discussed in the text.

A broadly similar trend is observed for the Σ5(310) GB with interstitially segregated P (Fig. 3d). Only one weak negative interaction, $J_{ij}$ = −0.16 mRy, is found between atoms lying within the GB plane; because this particular bond is oriented parallel to the viewing direction, it does not appear in the corresponding color map in Fig. 2c. The presence of interstitial P leads to an overall increase in Fe–Fe distances around the GB (see Fig. 3e), which significantly weakens interactions (I) and (IV) to approximately 0.2 mRy, although they remain positive. Interactions (II) and (III) are not evaluated because their interatomic distances exceed the threshold used in this study, becoming larger than even the third-shell distance in bulk Fe (7.65 a.u.), and thus fall outside the cutoff of 7.2 a.u. The interatomic distance associated with interaction (V) also becomes slightly larger than in the clean GB, but the corresponding exchange coupling is enhanced, reaching $J_{ij}$ = 1.69 mRy. Even stronger enhancements appear somewhat farther from the GB plane: nearest-neighbor interactions between atoms in the first and second, as well as the first and third adjacent layers, yield $J_{ij}$ = 2.22 mRy—almost twice as large as the bulk nearest-neighbor value $J_{NN}$.

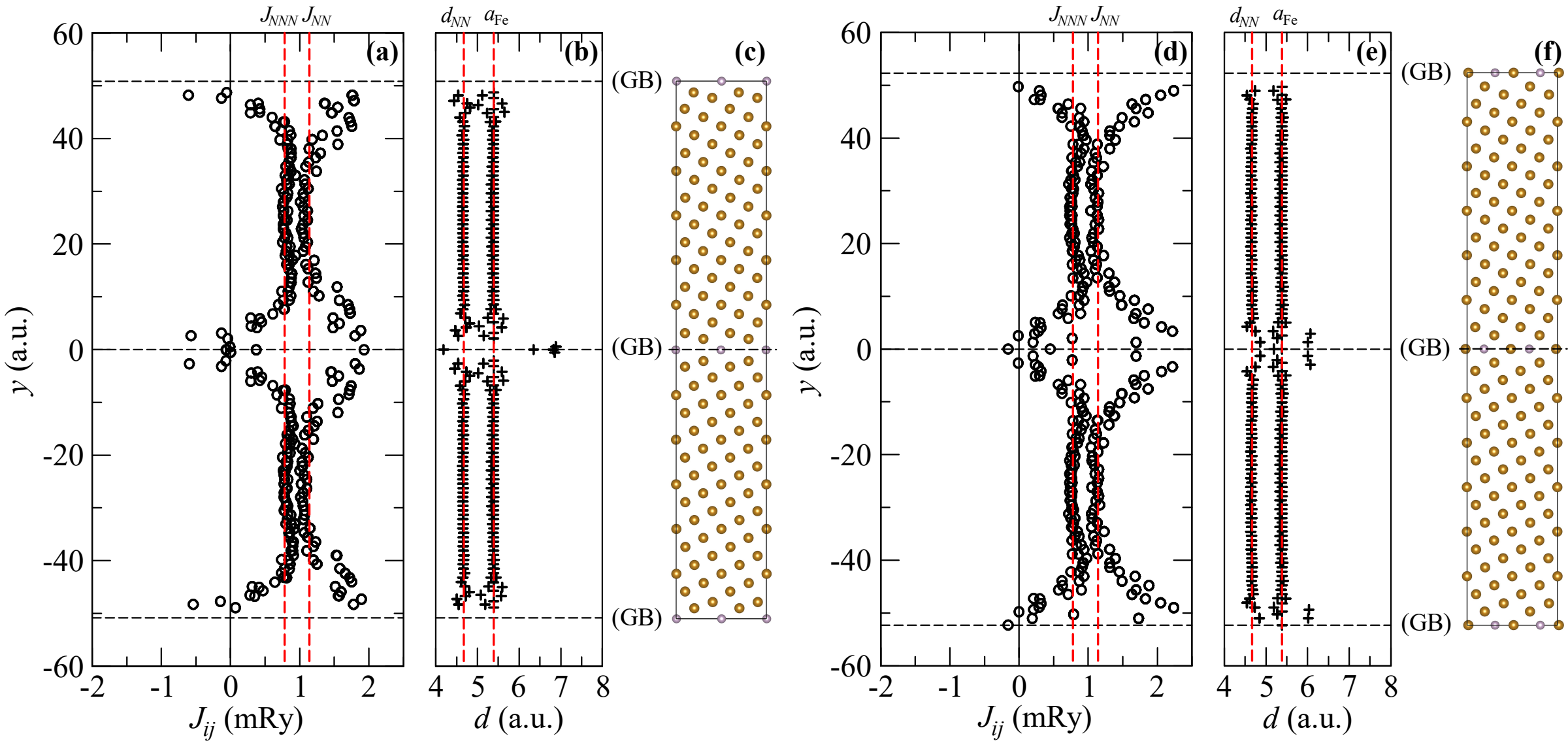

Fig. 3: (a,d) Exchange interaction parameters $J_{ij}$ and (b,e) corresponding interatomic distances *d* for the nearest and next-nearest neighbors in Σ5(310) GB with (a,b,c) substitutional and with (d,e,f) interstitial P at the GB plane. The *y*-axis represents the *y* coordinate of the center of the atomic pair *i*-*j* with respect to the center of the computational cell. The black dashed horizontal lines mark the GB planes, while the red dashed vertical lines denote the corresponding values of $J_{NN}$, $J_{NNN}$, $d_{NN}$ and $d_{NNN} = a_{\mathrm{Fe}}$ in bulk iron. Computational cells of Σ5(310) GBs with (c) substitutional and (f) interstitial P used for ab initio calculations.

As the clean Σ5(310) GB exhibits a strong antiferromagnetic exchange interaction across the GB plane, whereas the same boundary with segregated P impurity does not, we also examined two additional clean boundaries—Σ13(510) and Σ13(320)—to determine whether negative $J_{ij}$ values are a specific feature of the Σ5(310) atomic structure or a more general characteristic of similar GBs. The corresponding color maps for Σ13(510) and Σ13(320) are shown in Fig. 4a and 4b, respectively. Similar to the Σ5(310) GB, the Σ13(510) boundary also exhibits antiferromagnetic coupling between atoms in the first adjacent layers ((I), $J_{ij}$ = −1.02 mRy) and between atoms in the first and second adjacent layers ((II), $J_{ij}$ −0.43 mRy). These interactions correspond to the shortest (*d* = 4.24 a.u.) and the longest (*d* = 7.23 a.u.) interatomic distances examined in this supercell. However, the density of such antiferromagnetic pairs per unit GB area is smaller than in the Σ5(310) case. The strongest ferromagnetic interaction ($J_{ij}$ = 1.52 mRy, *d* = 4.53 a.u.) occurs between atoms located in the GB plane and atoms in the second adjacent layers (VII). The corresponding plots of $J_{ij}$ and the interatomic distances as a function of the *y*-coordinate of the pair center are shown in Fig. 5a and Fig. 5b, respectively. The spatial extent of the region influenced by the GB is similar to that of the Σ5(310) boundary, approximately 10 a.u. Enhanced nearest-neighbor interactions are again observed near the boundary, while the suppression of next-nearest-neighbor interactions appears more pronounced—likely due to the larger deviations of interatomic distances from the bulk values within the GB region.

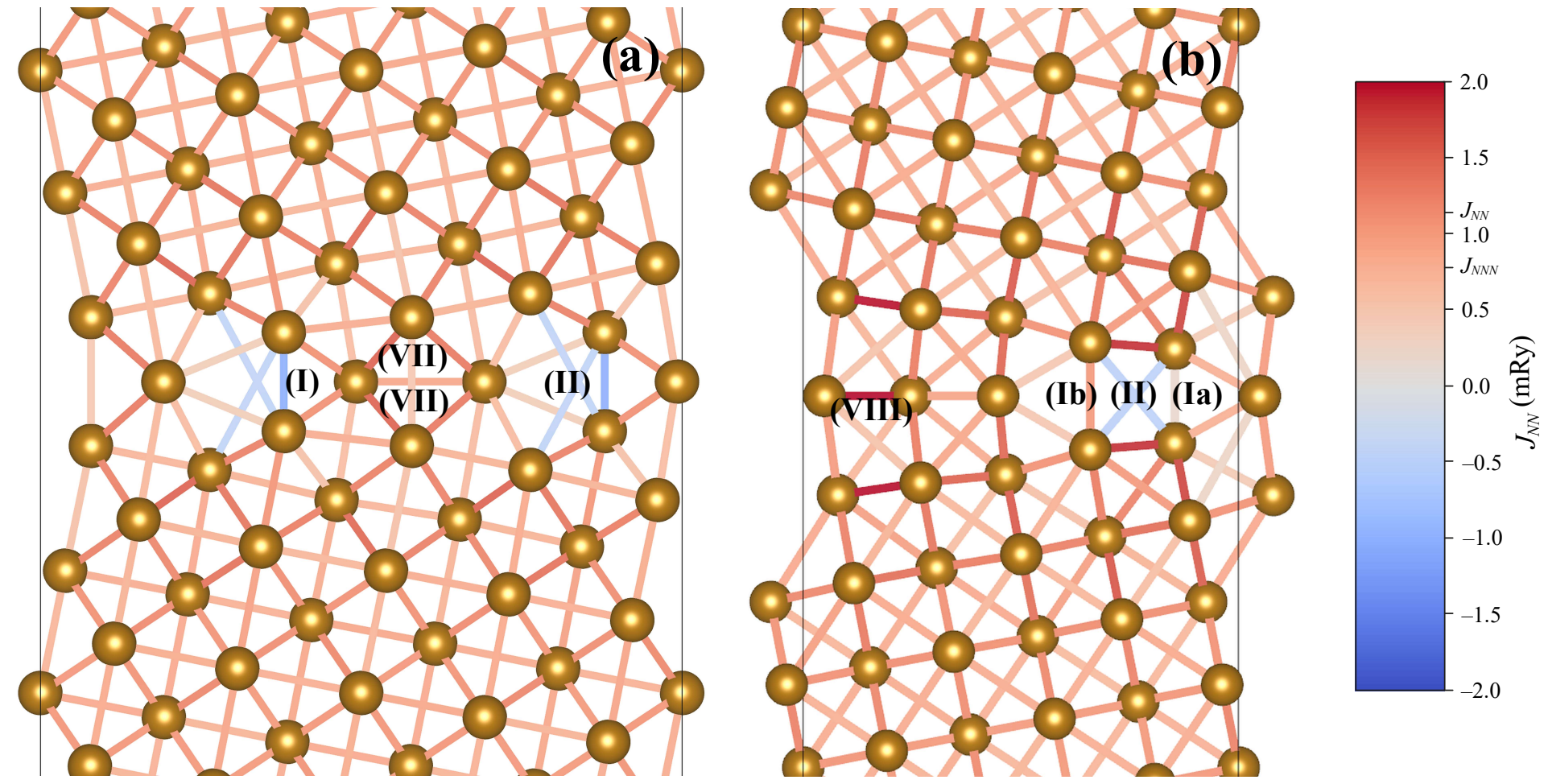

Fig. 4: Color maps of the exchange-interaction strength $J_{ij}$ for the clean (a) Σ13(510) and (b) Σ13(320) GBs. Color codes and labels are the same as in Fig. 2.

The Σ13(320) GB shows a somewhat different behavior, reflecting its distinct atomic structure near the boundary plane (Fig. 4b). Nevertheless, negative interactions are still present: a next-nearest-neighbor pair in the first adjacent layers (II) exhibits $J_{ij}$ = −0.47 mRy, corresponding to one of the longest interatomic distances analyzed, *d* = 6.47 a.u. Conversely, the shortest distance in this supercell, *d* = 4.21 a.u., occurs between nearest-neighbor atoms in the first layers across the GB (Ia), for which $J_{ij}$ is close to zero. A structurally similar interaction across the GB (Ib), with a slightly longer distance of *d* = 4.79 a.u., shows a positive coupling of $J_{ij}$ = 1.02 mRy, comparable to the bulk $J_{NN}$ value. The strongest exchange interaction, $J_{ij}$ = 2.36 mRy, is found between atoms lying directly in the GB plane (VIII), oriented perpendicular to the viewing direction; the corresponding interatomic distance is 4.70 a.u. Both enhancement and suppression of interaction strengths relative to bulk values are considerably more pronounced than in the Σ5(310) and Σ13(510) cases, but occur over approximately the same spatial region around the GB plane, as shown in Fig. 5(d).

To further examine how the antiferromagnetic coupling across the GB plane affects the magnetic properties of iron, we performed MC simulations of the temperature dependence of the magnetization, the susceptibility and the heat capacity (Fig. 6). Among the studied boundaries, only the Σ5(310) GB was selected for this analysis, as it exhibits the highest density of interactions with negative $J_{ij}$. Figure 6a shows the magnetization as a function of temperature for a supercell containing the Σ5(310) GB at the same GB–GB spacing ($d_{GB}$ = 51.2 a.u.) as in the *ab initio* calculations, together with the corresponding curve for defect-free bulk iron. The curve with GB is slightly shifted towards lower temperature, indicating a small decrease in the Curie temperature, as clearly confirmed by the shift of the susceptibility peak corresponding to the Curie point (Fig. 6b).

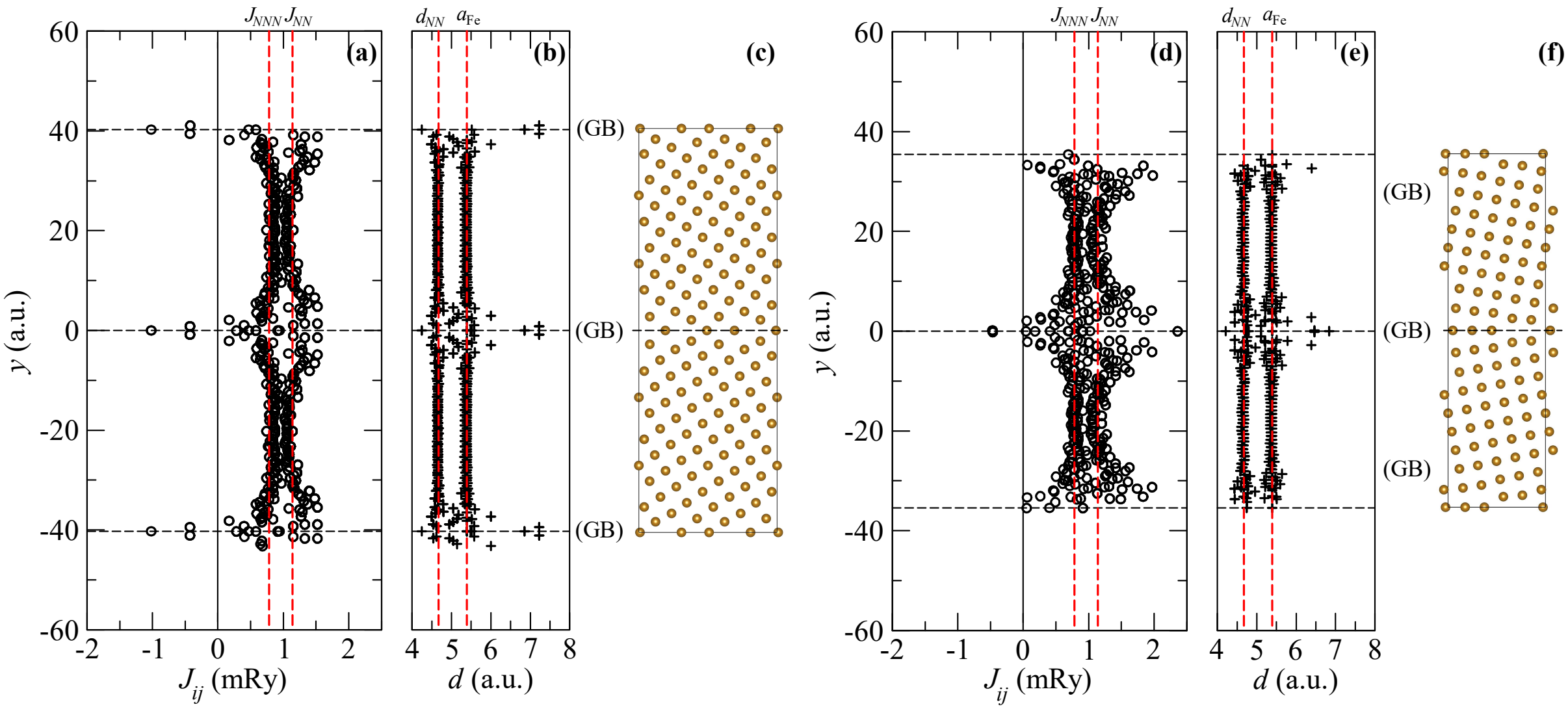

Fig. 5: (a,d) Exchange interaction parameters $J_{ij}$ and (b,e) corresponding interatomic distances *d* for the nearest and next-nearest neighbors in Σ13(510) (a,b,c) and Σ13(320) (d,e,f) GBs. The *y*-axis represents the y coordinate of the center of the atomic pair *i-j* with respect to the center of the computational cell. The black dashed horizontal lines mark the GB planes, while the red dashed vertical lines denote the corresponding values of $J_{NN}$, $J_{NNN}$, $d_{NN}$ and $d_{NNN} = a_{\mathrm{Fe}}$ in bulk iron. Computational cells of Σ13(510) (c) and Σ13(320) (f) GBs used for ab initio calculations.

To evaluate how decreasing grain size (i.e., increasing GB density) influences the magnetic transition, we artificially reduced the GB–GB spacing by removing several atomic planes, obtaining a separation of 17.1 a.u. (Fig. 6d). The exchange interactions between the remaining atoms were kept identical to those shown in Fig. 2a. We note that this atomic configuration does not correspond to a realistic grain structure; it is introduced solely to assess the effect of reducing the distance between GBs. This modified geometry results in a noticeable reduction in $T_C$ by approximately 100 K (Fig 6a-c). Overall, the results indicate that the influence of the Σ5(310) GB on the Curie temperature is small, but becomes significant when the GB density is substantially increased.

**Discussion**

Comparison of the three clean GBs shown in Fig. 2a and Fig. 4a–b reveals that antiferromagnetic exchange interactions across the GB plane consistently appear whenever the local atomic environment deviates substantially from the bulk coordination. The density of such configurations strongly depends on the GB tilt angle: the Σ5(310) GB exhibits the highest density of negative $J_{ij}$ values, whereas the Σ13(320) GB shows the lowest ones. To further demonstrate that the negative exchange interactions originate primarily from the broken local symmetry and the modified coordination of atoms near the boundary, rather than from changes in interatomic distance alone, we plotted the $J_{ij}$ parameters for all three GBs as a function of interatomic distance (Fig. 7). For comparison, we also included the nearest-neighbor ($J_{NN}$) and next-nearest-neighbor ($J_{NNN}$) exchange interactions of bulk bcc Fe calculated at different lattice constants *a*. The bulk trends agree well with previously published results [39]: $J_{NN}$ increases with atomic spacing up to approximately 1.8 mRy, then slightly decreases beyond the equilibrium distance before exhibiting a small increase and becoming nearly constant. $J_{NNN}$, on the other hand, increases up to about 1.0 mRy just beyond the equilibrium spacing and then gradually decreases. By comparison, the values obtained for atoms near the GBs show no such systematic dependence. Importantly, negative $J_{ij}$ values occur for both the shortest and the longest interatomic distances, underlining that distance alone cannot account for the observed behavior. Instead, the wide spread of $J_{ij}$ values, and especially the negative interactions, reflects the dominant influence of the local coordination geometry and symmetry breaking characteristic of the GB environment.

The results for the Σ5(310) GB with segregated P confirm that chemical perturbations can strongly alter the magnetic signature of a grain boundary. Phosphorus suppresses the antiferromagnetic interactions observed across the clean boundary and redistributes the pattern of enhanced and weakened couplings over a broader region. To further separate the chemical effect of P from the effect of P-induced structural relaxation, we performed additional calculations, summarized in Fig. 8. First, $J_{ij}$ was calculated for the relaxed Σ5(310) GB shown in Fig.2(a) but with P atoms additionally inserted into interstitial positions, without allowing the Fe lattice to relax further. In this configuration, the exchange interactions are not affected by the expansion of interatomic distances that appears after full structural relaxation with P atoms. The corresponding color map is shown in Fig. 8(a). Comparison with Fig. 2(a) shows that the presence of P changes interaction (I) across the GB plane from antiferromagnetic to ferromagnetic. Interestingly, interaction (III) across the GB plane changes in the opposite direction, from ferromagnetic to antiferromagnetic. In addition, P enhances several other interactions in the vicinity of the GB, in particular between atoms at the GB and atoms in the second layer, and between atoms in the first and second layers. This demonstrates the dominant role of the local electronic and chemical environment introduced by P.

We also calculated $J_{ij}$ for the fully relaxed Σ5(310) GB with interstitial P (see Fig.2(c)) after removing the P atoms from the relaxed structure. This allows the effect of P-induced lattice distortion to be separated from the direct chemical effect of P. Comparison of Fig. 8(b) with Fig. 2(c) shows that interactions (I) and (IV) in the vicinity of the P atoms are slightly enhanced when P is present, but decrease nearly to zero after P is removed. This further confirms the importance of the local chemical environment. By contrast, interactions farther from the original P positions are not significantly affected by removing P, suggesting that lattice distortion plays a more important role in those regions.

Finally, because the high local P concentration in the original model may influence the results, we also doubled the supercell in the z-direction and introduced only a single P atom in the GB plane, reducing the P concentration by a factor of four; the corresponding results are shown in Fig. 8(c). The antiferromagnetic interaction across the GB plane disappears also in this lower-concentration configuration, including in the part of the GB where P is not directly present. This confirms the strong influence of the local electronic and chemical environment introduced by P, while also indicating a secondary but still important contribution from the accompanying lattice distortion.

The discussion above focuses on the interstitial P configuration, which was identified as the more favorable configuration in the present model [43,51]. The configuration with substitutional P should therefore be interpreted with caution. In contrast to interstitial segregation, substitution of an entire atomic plane by P represents an extreme local concentration. This configuration is better viewed as a limiting case resembling the formation of a P-rich intergranular layer, rather than as a realistic dilute segregation state at a Fe grain boundary. Consequently, the corresponding changes in $J_{ij}$ should be understood as the magnetic response to a strongly perturbed local chemical environment, not as a quantitatively transferable prediction for dilute substitutional P segregation.

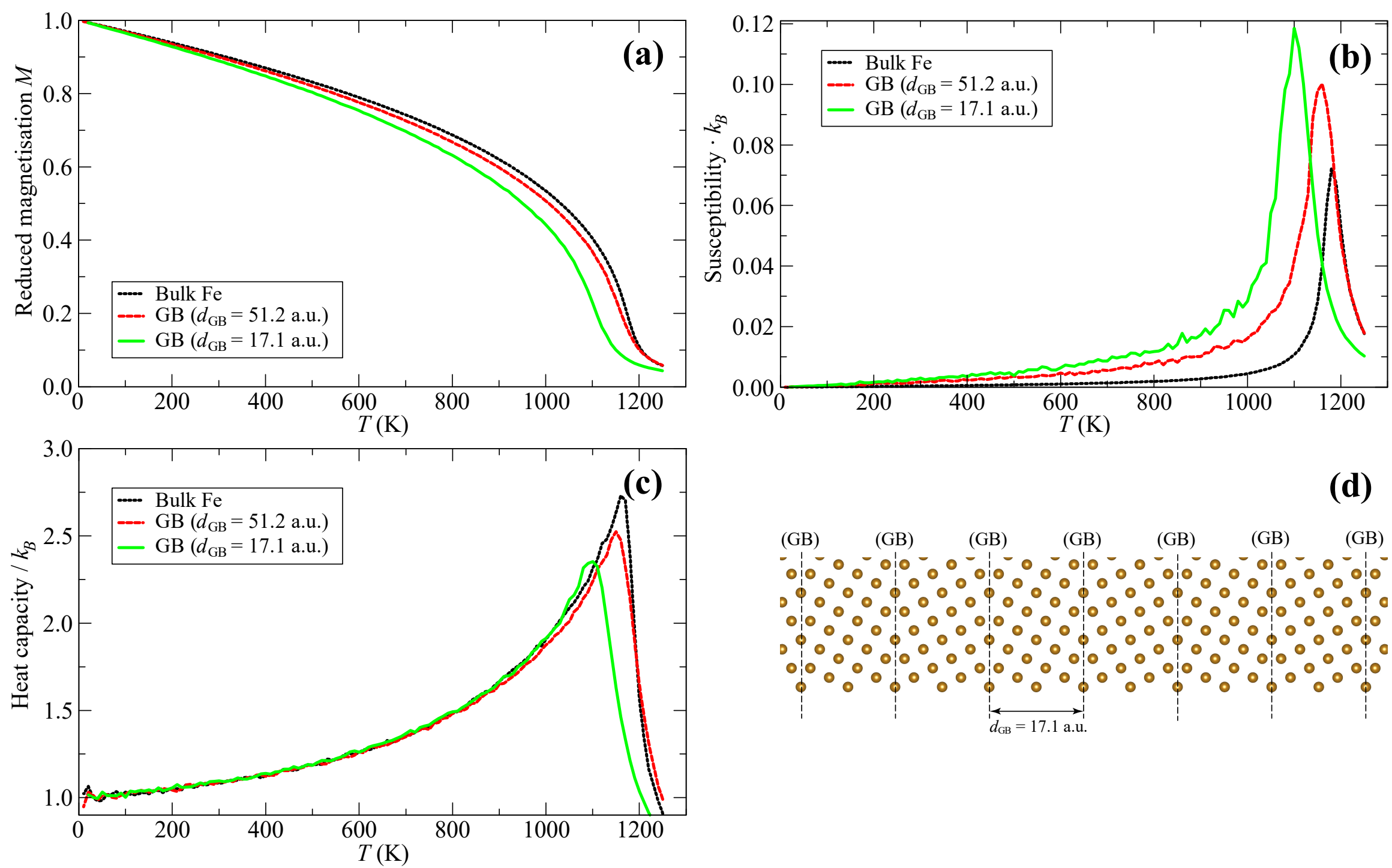

Fig. 6: (a) Magnetization, (b) susceptibility and (c) heat capacity as a function of temperature, calculated by MC simulations for bulk Fe and for two different Σ5(310) GB densities (i.e., two different values of $d_{GB}$). (d) Structural model used for the MC simulation in which the GB density was artificially increased compared to the structural model used in the ab initio calculations (see Fig. 1(c)).

The Monte Carlo simulations based on calculated exchange parameters provide a complementary, mesoscopic perspective on the role of GBs. Despite the presence of pronounced local antiferromagnetic couplings across the Σ5(310) GB and the strong spatial modulation of $J_{ij}$ near the boundary, the Curie temperature slightly decreases when the GB spacing corresponds to that used in the DFT supercell ($d_{GB}$ = 51.2 a.u.). Only when the GB–GB distance is artificially reduced to 17.1 a.u.—thereby significantly increasing the volume fraction affected by the GB structure—does $T_C$ decrease noticeably by about 100 K. This behavior demonstrates that, for realistic GB densities, the impact of even strongly modified local exchange interactions on the global magnetic ordering temperature is relatively weak.

This apparent robustness of $T_C$ can be rationalized in terms of the spatial distribution and mutual compensation of exchange interactions near the GB. The antiferromagnetic couplings across the boundary are confined to a narrow region and coexist with enhanced ferromagnetic interactions in neighboring shells. As a result, the net effect on the average exchange field that governs the onset of long-range ferromagnetism is small, even though the local magnetic environment near the GB is highly inhomogeneous. In other words, GBs primarily introduce local frustration and short-range magnetic disorder, while the bulk-like regions between them continue to dominate the global thermodynamic transition.

At the same time, the strong local deviations of $J_{ij}$ around GBs—especially the presence of antiferromagnetic bonds and strongly enhanced ferromagnetic couplings—are expected to have a much more pronounced influence on properties that are sensitive to local magnetic structure, such as domain-wall pinning, coercivity, and magnetization reversal processes. Thus, although the effect of isolated GBs on $T_C$ in Fe-like systems appears limited, their local modification of exchange interactions may still be important for micromagnetic behavior. Extending the present framework to include anisotropy, magnetoelastic coupling, and more realistic GB networks in polycrystalline microstructures will be an important next step towards linking atomistically resolved $J_{ij}$ to coercivity and loss behavior in technologically relevant soft magnetic materials.

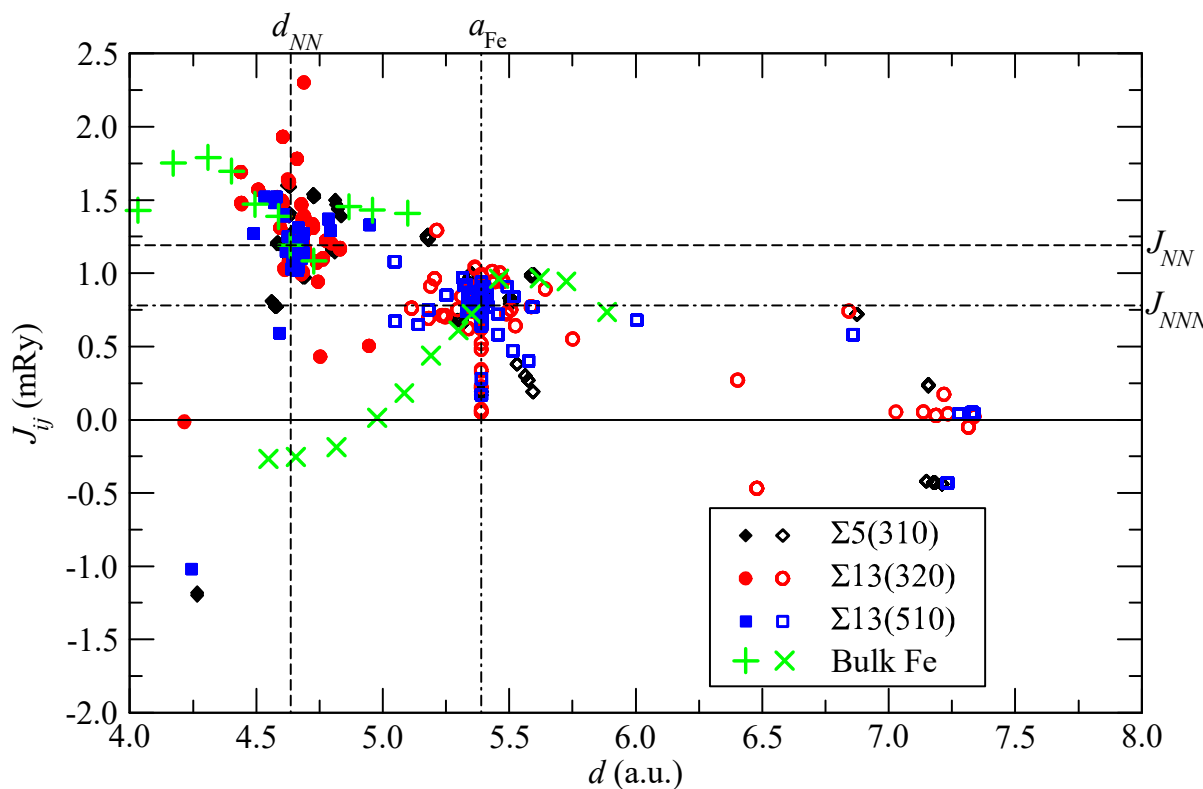

Fig. 7: Exchange interaction parameters $J_{ij}$ as a function of interatomic distances *d* for bulk Fe and all three investigated clean GBs. Filled and plus (+) symbols correspond to data for nearest neighbors, whereas open and cross (×) symbols denote data for next-nearest neighbors. The cross (×) symbols then show the variation in the lattice constant *a*.

The present methodology can be transferred from bcc Fe to more application-oriented magnetic materials in which GBs and intergranular regions play a central role. In Fe-based nanocrystalline soft magnets, such as FINEMET-type Fe–Si–B–Nb–Cu alloys, excellent soft-magnetic behavior is achieved through ultrafine bcc Fe-based grains, and the coercivity and permeability are strongly controlled by grain size, exchange averaging over many grains, and the magnetic character of the residual boundary phase [25,52]. In rare-earth permanent magnets, especially Nd–Fe–B-based systems, coercivity is likewise highly sensitive to grain-boundary structure and chemistry; grain-boundary diffusion or infiltration with rare-earth-rich phases is widely used to modify the effective magnetic coupling between neighboring grains and suppress magnetization reversal, while reducing the amount of heavy rare-earth elements required for high coercivity [53,54]. Atomistically resolved $J_{ij}$ calculations do not directly equal such coarse-grained intergranular coupling parameters, but they provide microscopic exchange information from which effective interface or grain-boundary coupling parameters can be derived or constrained in atomistic spin and micromagnetic models. To the best of our knowledge, the present work provides the first such $J_{ij}$ dataset for GBs in bcc Fe, and thus establishes a test case for extending the same framework to chemically complex soft and hard magnetic materials.

The selected symmetric GBs should therefore be viewed as representative model interfaces rather than as a statistically complete description of all GBs in Fe. Their role in the present work is to provide controlled local environments in which the effect of boundary structure and segregation on atomistic exchange interactions can be quantified. Generalization to real polycrystalline materials would require sampling a wider grain-boundary character distribution, including different misorientations, boundary planes, segregation configurations, and local chemical environments. Nevertheless, the same sequence of atomistic structure generation, first-principles electronic-structure calculation, extraction of $J_{ij}$ parameters, and finite-temperature magnetic simulation can be applied to these cases once the corresponding atomic structures are available.

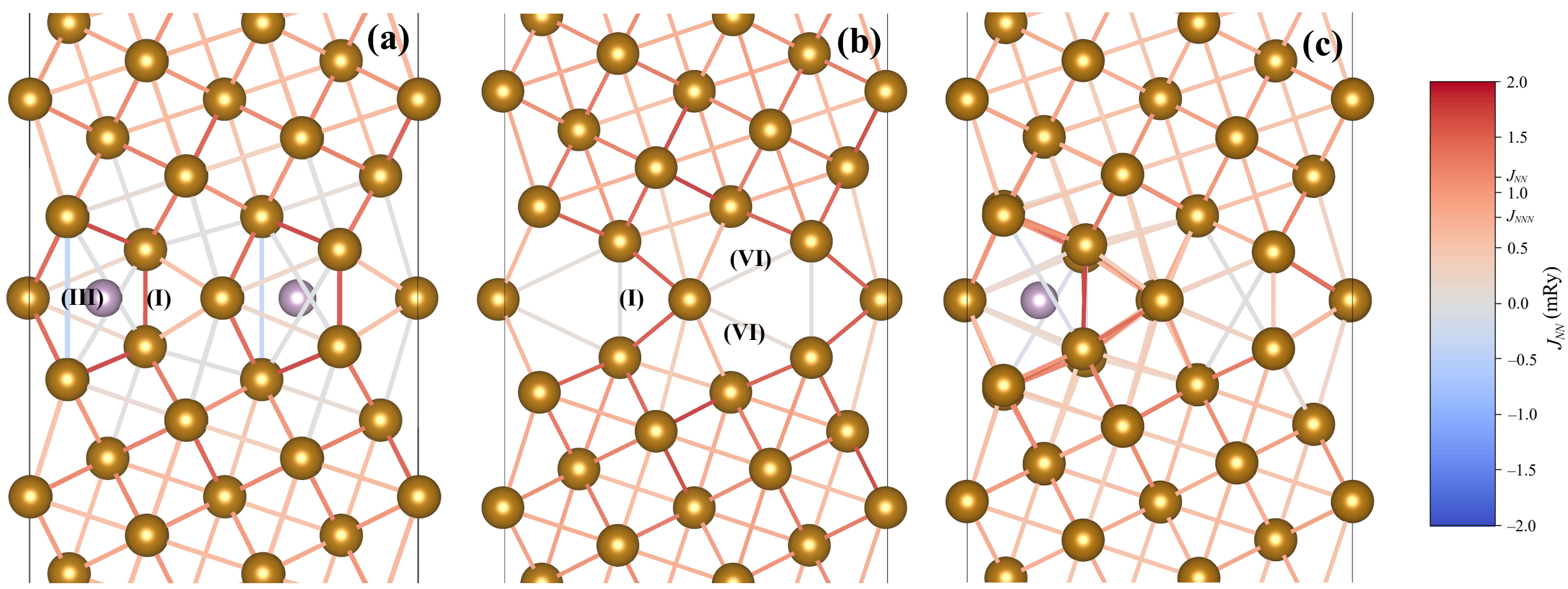

Fig. 8: Color maps of the exchange-interaction strength $J_{ij}$ for (a) the relaxed clean Σ5(310) GB with interstitial P atoms inserted without further structural relaxation, (b) the relaxed GB with interstitial P after subsequent removal of the P atoms, and (c) the relaxed GB with the P concentration reduced by a factor of four. Color codes and interaction labels are the same as in Fig. 2.

Extending this strategy from selected model GBs to realistic polycrystalline microstructures would require sampling many more boundary characters, segregation patterns, local strains, and chemical environments. Since fully first-principles $J_{ij}$ calculations for all such cases may become computationally demanding, machine-learning (ML) approaches provide one possible route to accelerate this step and to apply the present framework to more complex structures. The exchange parameters calculated in this work may therefore serve as reference data for ML descriptions of magnetic materials. The present work extends previous first-principles studies of Fe GB magnetism beyond local magnetic moments and electronic-structure changes. This is relevant because several magnetic ML interatomic potentials have already been developed specifically for Fe or Fe-based systems, including magnetic Moment Tensor Potentials (MTP) for collinear spin-polarized bcc Fe [55], ML spin-lattice potentials for defective magnetic Fe [56], and non-collinear magnetic Atomic Cluster Expansion (ACE) models for Fe [57]. These methods aim to describe the coupling between atomic structure and magnetic degrees of freedom, and grain boundaries represent a particularly demanding test case because they combine strong local structural distortion, reduced symmetry, and modified magnetic coordination.

In this context, the present grain-boundary $J_{ij}$ dataset could be used in several complementary ways. Models that explicitly represent Heisenberg-type pair interactions, such as spin-dependent graph neural-network potentials with Heisenberg-edge terms [58], could in principle be trained or benchmarked directly against the calculated $J_{ij}$ values. More general magnetic ML potentials, such as magnetic MTP, magnetic ACE, or spin-lattice neural-network potentials, are usually trained on total energies, forces, magnetic moments, or spin configurations; for these models, effective $J_{ij}$ values could be obtained a posteriori by mapping energies of different magnetic configurations onto a Heisenberg Hamiltonian [17], by evaluating energy changes under small rotations of selected magnetic moments, or by fitting the spin-dependent part of the learned energy surface to pairwise exchange terms. Comparison with the present DFT- LKAG results would then test whether the ML model reproduces not only structural energetics, but also the sign, magnitude, distance dependence, and dominant exchange paths of the magnetic interactions governing spin ordering near Fe grain boundaries.

## Conclusions

In the present work, we employed density-functional theory in combination with a tight-binding Green's-function approach to determine the magnetic exchange-interaction parameters $J_{ij}$ at grain boundaries in bcc Fe. All three studied boundaries—Σ5(310), Σ13(510), and Σ13(320)—exhibit antiferromagnetic coupling between atoms across the GB plane. These interactions occur whenever the local atomic environment deviates substantially from the bulk coordination and are not governed by interatomic distance alone. In contrast, phosphorus segregation, whether interstitial or substitutional, suppresses this antiferromagnetic behavior and significantly reshapes the local exchange landscape through chemical and electronic effects. Monte Carlo simulations of magnetization, susceptibility and heat capacity further show that the bulk-like regions dominate the global thermodynamic transition to the paramagnetic state, while the narrow GB region contributes only weakly. However, a very large increase in GB volume fraction leads to a substantial reduction in the Curie temperature.

Overall, our results highlight that grain boundaries strongly modify local magnetic interactions but have a limited influence on global magnetic ordering in Fe. The methodology demonstrated here provides a robust framework for linking atomistic variations of $J_{ij}$ at interfaces to mesoscale magnetic behavior and can be directly extended to more complex alloy systems and technologically relevant soft magnetic materials.

## Data availability statement

All data that support the findings of this study are included within the article (and any supplementary files).

## Acknowledgements

This work was financially supported by the Ministry of Education, Youth and Sports of the Czech Republic (Project No. LUC25051) and by Brno University of Technology (Project No. FSI-S-23-8225). Computational resources were provided under the Projects e-INFRA CZ (ID:90254) at the IT4Innovations National Supercomputing Center. The authors also acknowledge the "Centre Régional Informatique et d'Applications Numériques de Normandie" (CRIANN) where simulations were performed as Project No. 2015004. Figures were visualized using the VESTA software [59] (version 3, National Museum of Nature and Science, 4-1-1, Amakubo, Tsukuba-shi, Ibaraki 305-0005, Japan).